\documentclass[preprint,3p, twocolumn]{elsarticle}
\usepackage[fleqn]{amsmath}
\usepackage{amssymb,amsthm}
\usepackage{mathtools}
\usepackage{mathrsfs}
\usepackage{bm}
\usepackage{slashed}	%feynman slashed symbols
\usepackage{cancel}
\usepackage{soul}
\usepackage{color}
\usepackage{xcolor}
\usepackage[pdftex,hypertexnames=false,linktocpage=true]{hyperref}
\hypersetup{colorlinks=true,linkcolor=red,anchorcolor=darkgreen,citecolor=green!50!blue,filecolor=darkgreen,urlcolor=green,
bookmarksnumbered=true,
pdfview=FitB
}
\usepackage{lineno}
\biboptions{sort&compress}
\usepackage{verbatim}
\usepackage{fullpage}
\usepackage{graphicx}
\usepackage{subfig} 	% a set of figures
\usepackage{relsize}	%rescalable math
\usepackage{float}
\usepackage{array}
\usepackage{booktabs}
\usepackage{arydshln}
\usepackage{multirow}
\makeatletter

\newcommand{\Rmnum}[1]{\expandafter\@slowromancap\romannumeral #1@}
\makeatother
\colorlet{darkgreen}{green!60!black}
\colorlet{brightyellow}{yellow!75!red}
\colorlet{orange}{red!50!yellow}
\colorlet{darkblue}{blue!60!black}
\colorlet{darkred}{red!80!black}
\colorlet{greenblue}{green!50!blue}

\newcommand{\be}{\begin{eqnarray}}
\newcommand{\ee}{\end{eqnarray}}
\newcommand{\eq}{\begin{eqnarray}}
\newcommand{\en}{\end{eqnarray}}

\newcommand{\nn}{\nonumber}

\colorlet{purple1}{blue!70!red}
\colorlet{darkred}{red!50!black}

\usepackage{color}
\usepackage{xcolor}
\usepackage{soul}
\usepackage{verbatim} %%%% for commenting the lines
\usepackage{siunitx}
\usepackage{hyperref}
\hypersetup{colorlinks=true,linkcolor=purple,anchorcolor=blue,citecolor=purple1, filecolor=blue,urlcolor=red,bookmarksnumbered=true,
pdfview=FitB
}

 \journal{Physics Letters B}

\begin{document}
\begin{frontmatter}

%title and authors
\title{Basis light-front quantization approach to photon}

\author[imp,cas,impcas]{Sreeraj Nair}
\ead{sreeraj@impcas.ac.cn}
\author[imp,cas,impcas]{Chandan Mondal}%\corref{c1}}
\ead{mondal@impcas.ac.cn}
\author[imp,cas,impcas]{Xingbo Zhao}  
\ead{xbzhao@impcas.ac.cn}
\author[iitb]{Asmita Mukherjee}  
\ead{asmita@phy.iitb.ac.in}
\author[ISU]{James P. Vary}  
\ead{jvary@iastate.edu}

%\author[]{(BLFQ Collaboration)}

\author[]{\\\vspace{0.2cm}(BLFQ Collaboration)}

%\cortext[c1]{Corresponding author}

\address[imp]{Institute for Modern Physics, Chinese Academy of Sciences, Lanzhou-730000, China} 
\address[cas]{School of Nuclear Science and Technology, University of Chinese Academy of Sciences, Beijing 100049, China}
\address[impcas]{CAS Key Laboratory of High Precision Nuclear Spectroscopy, Institute of Modern Physics, Chinese Academy of Sciences, Lanzhou 730000, China}
\address[iitb]{Department of Physics, Indian Institute of Technology Bombay, Powai, Mumbai 400076, India}
\address[ISU]{Department of Physics and Astronomy, Iowa State University, Ames, Iowa 50011, USA}
%%abstract

\begin{abstract}
We solve for the light-front wave functions (LFWFs) of the physical photon from the eigenvectors of the light-front quantum electrodynamics (QED) Hamiltonian with the aim to determine its bare photon and electron-positron Fock components.  We then employ the resulting LFWFs to compute the transverse momentum dependent parton distributions (TMDs) and the generalized parton distributions (GPDs) of the photon. The TMDs are found to be in excellent agreement with the lowest-order perturbative results calculated using the electron-positron quantum fluctuation of the photon. The GPDs are also consistent with the perturbative calculations.

\end{abstract}
\begin{keyword}
Light-front quantization \sep Quantum electrodynamics \sep Photons \sep Parton distribution functions 
\end{keyword}
\end{frontmatter}

%{\it Introduction.}---
%======================
\section{Introduction}
%======================
%An essential tool for studying composite systems is the deep inelastic scattering (DIS) process, where individual partons are resolved. It is now well known that the photon has a partonic substructure, induced by its quantum fluctuations into charged fermion antifermion pairs. The parton content of the photon is known to play an important role in high energy scattering processes. The photon structure is probed from the DIS of a highly virtual photon on a real photon~\cite{Walsh:1973mz,Walsh:1972dp,Witten:1977ju,Berger:1981bh,Peterson:1982tt,Nisius:1999cv,Vogt:2005dw}. One can extract the structure functions or the parton distribution functions (PDFs) from such process, encoding the distribution of longitudinal momenta and polarizations carried by the partons.  However, they do not provide knowledge about the transverse motion and the spatial location of the partons inside the system. Meanwhile, it has become clear that both the transverse momentum dependent parton distributions (TMDs), appearing in the description of semi-inclusive processes~\cite{Mulders:1995dh,Barone:2001sp,Bacchetta:2006tn}, and the generalized parton distributions (GPDs), appearing in the description of hard exclusive reactions~\cite{Ji:1996nm,Diehl:2003ny,Belitsky:2005qn,Goeke:2001tz} provide us with essential information about the distribution and the orbital motion of partons inside composite systems, and allow us to draw three-dimensional pictures of the systems.

The unique feature of the photon is that its partonic content can be evaluated in the leading order perturbative QED~\cite{Friot:2006mm,ElBeiyad:2008ss}.
While the parton distributions (PDFs) of the photon are now well understood both theoretically and experimentally~\cite{Walsh:1973mz,Walsh:1972dp,Witten:1977ju,Berger:1981bh,Peterson:1982tt,Nisius:1999cv,Vogt:2005dw,Berger:2014rva}, much less information is available for its GPDs and TMDs. They provide us with essential information about the distribution and the orbital motion of partons inside composite systems as well as three-dimensional pictures of the systems~\cite{Ji:1996nm,Diehl:2003ny,Belitsky:2005qn,Goeke:2001tz,Mulders:1995dh,Barone:2001sp,Bacchetta:2006tn}.

The photon GPDs were first introduced in Ref.~\cite{Friot:2006mm}, where  the authors considered the deeply virtual Compton scattering (DVCS), $\gamma^* \gamma \rightarrow \gamma \gamma$, assuming the nearly real photon as a photon target. Later, the photon GPDs have been studied using the light-front wave function of the photon obtained analytically using perturbation theory~\cite{Mukherjee:2011bn,Mukherjee:2011an,Mukherjee:2013yf}. The analytic properties of DVCS amplitudes and related sum rules of the photon GPDs have been investigated in Ref.~\cite{Gabdrakhmanov:2012aa}. 

The TMDs are the extended version of collinear PDFs, predicting the three-dimensional structural information of the composite system in momentum space. These distributions also help to gain the knowledge about the correlation between spins of the target and the partons.  The conventionally used methods for measuring TMDs are semi-inclusive deep inelastic scattering~\cite{Brodsky:2002cx,Ji:2004wu,Bacchetta:2017gcc} and the Drell-Yan process~\cite{Ralston:1979ys,Tangerman:1994eh,Collins:2002kn,Zhou:2009jm}. 
%Unlike the spin-$\frac{1}{2}$ hadron such as nucleon, the spin-$1$ systems, like deuteron or vector mesons have more TMDs following the increase in the spin degrees of freedom. 
 While the TMDs of hadrons have been widely investigated both experimentally and theoretically, the TMDs for the photon remain unexplored. Meanwhile, the TMDs of a spin-$1$ target in quantum chromodynamics (QCD) have been studied in Refs.~\cite{Bacchetta:2001rb,Ninomiya:2017ggn,Kaur:2020emh,Kumano:2020ijt}. In this work, we investigate both the GPDs and the TMDs of the physical spin-$1$ particle in QED, i.e., the photon within basis light-front quantization (BLFQ), which provides a nonperturbative framework for solving relativistic many-body bound state problems in quantum field theories~\cite{Vary:2009gt,Honkanen:2010rc,Zhao:2014xaa,Wiecki:2014ola,Li:2015zda,Jia:2018ary,Lan:2019vui,Qian:2020utg,Xu:2019xhk,Xu:2021wwj}. Previously, this approach has been successfully applied to explore the GPDs~\cite{Chakrabarti:2014cwa} and the TMDs~\cite{Hu:2020arv} of the physical spin-$\frac{1}{2}$ particle in QED, i.e., electron.
 
With the framework of BLFQ~\cite{Vary:2009gt}, we consider the light-front QED Hamiltonian and solve for its mass eigenvalues and eigenstates. With electron (positron) and photon being the explicit degrees of freedom for the electromagnetic interaction, our Hamiltonian incorporates light-front QED interactions relevant to constituent bare photon and electron-positron Fock components of the physical photon. By solving this Hamiltonian in the leading two Fock sectors and using the physical electron mass, i.e., $m_e=0.511$ MeV and the physical electromagnetic coupling constant, i.e., $e=0.3$, we obtain a good quality description of the photon TMDs and GPDs from our resulting LFWFs obtained as eigenvectors of this Hamiltonian. We then compare the BLFQ computations with the results evaluated using perturbative LFWFs for the photon~\cite{Harindranath:1998pd}. 
It is our aim to manifest that the BLFQ framework, which is innately nonperturbative, gives results, up to finite basis artifacts, that agree well with the
perturbative results.
%===================================================================================
\section{Photon wave functions with BLFQ} \label{secblfq}
%===================================================================================
The BLFQ aims at solving the following eigenvalue equation to obtain the mass spectra and the LFWFs of hadronic bound states:
\begin{align}
H_\mathrm{{LF}} \mid \Psi_{\rm h} \rangle = M^2_{\rm h} \mid \Psi_{\rm h} \rangle \,,
\label{eveq}
\end{align}
where $H_\mathrm{{LF}}  =  P^+ P^- - (P^{\perp})^2$ is the light-front Hamiltonian and the operators $P^+ $, $P^-$ and $P^{\perp}$ are the longitudinal momentum, the dynamical generator of translations in light-front time and the transverse momentum, respectively of the system.
The diagonalization of Eq.~(\ref{eveq}) using a suitable matrix representation for the Hamiltonian generates the eigenvalues $M^2_{\rm h}$ that correspond to the mass-squared spectrum and the associated eigenstates $\mid \Psi_{\rm h} \rangle$, which encode structural information of the bound states. 

Like hadrons in QCD, we treat the physical photon as a
composite particle, where we could find the bare photon, electrons, and
positrons as its partons. At fixed light-front time, the physical photon state can be expressed in terms of photon ($\gamma$), electron ($e$) and positron ($\bar{e}$), etc., Fock components,
\begin{align}\label{Eq1}
\mid \Psi_\gamma \rangle=\psi_{(\gamma)}\mid \gamma\rangle + \psi_{(e\bar{e})}\mid e\bar{e}\rangle+\dots\, , 
\end{align}
where the $\psi_{(\dots)}$ correspond to the probability amplitudes to find different parton configurations in the physical photon. In this work, we restrict ourselves to only the constituent photon and electron-positron Fock sectors and adopt the light-front QED $\hat{P}^{-}_{\mathrm{QED}}$ that incorporates interactions relevant to those leading two Fock components of the physical photon,
\begin{align}
&\hat{P}^{-}_{\mathrm{QED}} = \int {\rm d}x^- {\rm d}^2x^{\perp}\Big\{
 \frac{1}{2}\bar{\Psi}
\gamma^+ \frac{m_e^2 + (i\partial^{\perp})^2}{i\partial^+}
 \Psi\nn\\
&+\frac{1}{2}A^j\left[m_{0\gamma}^2+(i\partial^{\perp})^2 \right]A^j
  + e\,\bar{\Psi} \gamma^{\mu} \Psi A_{\mu}\Big\}\,,
\label{eff}
\end{align}
where $\Psi(x)$ and $A_{\mu}(x)$ are the fermion and the gauge boson fields, respectively,  %. $g$ represents the the physical electromagnetic coupling constant, 
and $\gamma^+ = \gamma^0 + \gamma^3$, where $\gamma^\mu$ are the Dirac matrices. The first and second terms in Eq.~(\ref{eff}) correspond to the kinetic energies of the electron and the photon with bare mass $m_e$ and $m_{\gamma 0}$, respectively and the third term represents their interaction with coupling $e$. In our current treatment, the instantaneous fermion interaction does not contribute, while the instantaneous photon interaction either contributes to overall renormalization factors or contains small-$x$ divergences that need to be canceled by explicit photon
exchange contributions from higher Fock-sectors.
%~\cite{Zhao:2014xaa,Tang:1991rc,Kaluza:1991kx}. 
 We thus neglect the instantaneous interactions. 

 Following a Fock sector-dependent renormalization procedure~\cite{Karmanov:2008br,Karmanov:2012aj}, we
introduce a mass counter-term, $m_{\rm ct}=m_{0\gamma}-m_{r\gamma}$, that represents the photon mass correction due to the quantum fluctuations to the higher Fock sector $| e\bar{e}\rangle$,  where $m_{r\gamma}$ is the renormalized photon mass. We adjust $m_{\rm ct}$ such that the ground state eigenvalue of the Hamiltonian becomes equal to the physical photon mass squared, which is necessarily zero for the real photon. Note that the mass counter-term is a function of the ultraviolet (UV) cutoff and it increases as the cutoff increases~\cite{Brodsky:1997de}. 

 The basis states in the BLFQ framework are expanded in terms of Fock state sectors, where each basis state has the longitudinal, the transverse and the spin degrees of freedom~\cite{Zhao:2014xaa}.
The longitudinal coordinate $x^-$ is confined to a box of length $2L$ with
 antiperiodic (periodic) boundary conditions for the fermions (bosons). Therefore, the longitudinal momentum of the particle is parameterized as $p^+ = 2 \pi k/L$, where the  longitudinal quantum number $k=\frac{1}{2},~\frac{3}{2},~\frac{5}{2},\dots$ for fermions, whereas for bosons $k=1,~2,~3,\dots$. We omit the zero mode for bosons. All many-body basis states have the same total longitudinal momentum $P^+ =\sum_ip^+_i$, where the sum runs over the particles in a particular basis state. We then express $P^+$ using a dimensionless quantity $K = \sum_i k_i$ such that $P^+ = 2 \pi K/L$. The longitudinal momentum fraction $x$ is now defined as $x_i = p_i^+/P^+ = k_i/K$. The variable $K$ can be seen as the ``resolution" in the longitudinal direction, and hence a resolution onto the parton distribution functions. When $K \rightarrow \infty$, we approach the continuum limit in the longitudinal direction.
 
  Two-dimensional harmonic-oscillator (2D-HO) modes are chosen as the basis states for the transverse degrees of freedom~\cite{Vary:2009gt,Zhao:2014xaa}.
The HO states, $\Phi_{nm}({\bf p}_\perp;b)$, are parameterized by $n$, $m$, and $b$, where $n$ and $m$ are the principal and orbital angular quantum number, respectively and $b$ represents its scale parameter.
The continuum limit in the transverse direction is dictated by the parameter $N_{\rm max} \rightarrow \infty$ where $N_{\mathrm{max}} \ge \sum_i (2n_i + |m_i|+1)$. The truncation parameter $N_{\rm{max}}$ controls the transverse momentum covered by the HO basis functions. The $N_{\rm max}$ truncation acts implicitly as the ultraviolet (UV) and infrared (IR) cutoffs. In momentum space, the UV cutoff $\Lambda_{\rm UV} \simeq b\sqrt{N_{\rm max}}$ and the IR cutoff $\lambda_{\rm IR} \simeq b/\sqrt{N_{\rm max}}$. 

For the spin degrees of freedom, the quantum number $\lambda$ is used to identify the helicity of the particle. Thus, each single-particle basis state
is designated using four quantum numbers, $\alpha_i=\{x_i,\,n_i,\,m_i,\,\lambda_i\}$. All many-body basis states are selected to have well defined values of the total angular momentum projection
$
M_J=\sum_i\left(m_i+\lambda_i\right).
$

%Upon diagonalization of the full light-front Hamiltonian define in Eq.~(\ref{eff}) within the BLFQ basis, we obtain the eigenvalues that correspond to the mass spectrum. We also produce the eigenstates that represent the LFWFs in the BLFQ basis encoding the structural information of the systems. 
The resulting LFWFs of the dressed photon in momentum space are then expressed as
%\begin{align}
%&\psi_{\lambda_e,\lambda_{\bar{e}}}^{\Lambda}\left(x_e,{\bf p}_e^{\perp},x_{\bar{e}},{\bf p}_{\bar{e}}^{\perp}\right) 
%\nn\\
%=&\sum_{n_e,m_e,n_{\bar e},m_{\bar e}} \Big[\psi\left( \alpha_e,\alpha_{\bar{e}}\right) \nn\\
%&\quad\times\, \phi_{n_e,m_e }
%\left( {\bf p}_e^{\perp};b \right)\phi_{n_{\bar{e}},m_{\bar{e}}}\left( {\bf p}_{\bar{e}}^{\perp};b \right)\Big]\,,
%\label{single}
%\end{align} 
\begin{align}
&\psi_{\{\lambda_i\}}^{\Lambda}\left(x_i,{\bf p}_i^{\perp}\right) 
\nn\\
=&\sum_{\{n_i,m_i\}} \Big[\psi\left( \{\alpha_i\}\right)\prod_{i=e,\bar{e}}\phi_{n_i,m_i }
\left( {\bf p}_i^{\perp};b \right)\Big]\,,
\label{single}
\end{align} 
with $\psi\left( \{\alpha_i\}\right) = \langle \{\alpha_i\} \,|\, \Psi_\gamma(P,\Lambda) \rangle$ being the components of the eigenvectors generated form the diagonalization of $\hat{P}^{-}_{\mathrm{QED}}$ given in Eq.~(\ref{eff}) within the BLFQ basis, where $|\, \Psi_\gamma(P,\Lambda) \rangle$ represents the photon state with the momentum $P$ and the helicity $\Lambda$. These single-particle wave functions given in Eq.~(\ref{single}) contain center-of-mass (CM) excitations, which need to be factorized out. A detailed discussion of the CM factorization within the BLFQ framework can be found in Refs.~\cite{Wiecki:2014ola,Xu:2021wwj,Hu:2020arv}. After factorizing out the CM excitations, we obtain LFWFs in the relative coordinates and employ them to compute the TMDs and the GPDs of the physical photon.% in the Sec.~\ref{secobs}.

Note that the act of truncating the Fock components results in the violation of the Ward-identity and hence, we introduce a rescaling factor to counteract this violation as discussed in Refs.~\cite{Zhao:2014xaa,Brodsky:2004cx,Chakrabarti:2014cwa,Hu:2020arv}.  The rescaled photon's observable $\mathcal{O_\mathrm{rs}}$ is then given by
\begin{align}
\mathcal{O_\mathrm{rs}} = \frac{\mathcal{O}}{Z_2}\,,
\label{rs}
\end{align}
where $\mathcal{O}$ denotes the naive observable calculated directly by using the LFWF from Eq.~(\ref{single}) and the rescaling factor $Z_2$ is the photon wave function renormalization factor, which in our truncation can be interpreted as the probability of finding a bare photon within a physical photon, 
\begin{align}
Z_2 = \sum_{|\gamma\rangle} \mid \langle \gamma \mid \gamma_{\mathrm{phys}} \rangle \mid^2\,,
\end{align}
where the summation runs over all the basis states in the $|\gamma\rangle$ sector. The  $Z_2$ incorporates the contribution from the quantum fluctuation between the $|\gamma\rangle$ and $|e\bar{e}\rangle$.
%===============================================================
%\begin{figure*}
%\begin{center}
%\includegraphics[width=0.45\linewidth]{sfcompared_exp_pluto_blfq.pdf}
%\includegraphics[width=0.45\linewidth]{sfcompared_combined.pdf}
%\caption{The photon structure function $F_{2,{\rm QED}}$ as a function of $x$. Left panel compares the BLFQ results with the experimental data~\cite{Nisius:1999cv}. The BLFQ results are evaluated with different $N_{\mathrm{max}}=\{18,\,31,\,53,\,222\}$, which correspond to different energy scales, $\mu_{\mathrm{N_{\mathrm{max}}}} = 2b^2N_{\mathrm{max}}\sim\{3.24,\,5.58,\,9.54,\,40\}$ GeV$^2$, corresponding to the energy scales of the experiments. Right panel compares the BLFQ results with the perturbative results evaluated with the UV cutoff $\Lambda_{\rm UV}\approx b\sqrt{x(1-x)2N_{\rm max}}$ employed in both the approaches. The results are obtained with three different  $N_{\mathrm{max}} = \{50,100,150\}$. The lines with dots represent the BLFQ computations, whereas the lines without dots correspond to the perturbative results.}
%\label{fig:F2}
%\end{center}
%\end{figure*}
%================================================================ 
%%==============================================
\section{Photon Observables} \label{secobs}
%==============================================
We treat the physical photon as a spin-$1$ composite particle, with a bare photon and a electron-positron pair that emerges from quantum fluctuation as its partons.  We employ our resulting LFWFs to compute the TMDs and the GPDs of the electron inside the physical photon. %and the structure function of the photon. 
%==============================================
\subsection{TMDs} \label{sec:TMDs}
%==============================================
Under the  approximation that the gauge link is the identity operator, the unpolarized and the polarized TMDs can be expressed in terms of the photon LFWFs as
\begin{align}
&f^{1}_{\gamma}(x,{\bf k}^{\perp 2}) = \frac{1}{2}\int [{\rm d}e\bar{e}]\nn\\&\times\sum_{\Lambda,\lambda_i}
\Psi_{\{\lambda_i\}}^{\Lambda*}\left(x_i,{\bf k}_i^{\perp}\right)
\Psi_{\{\lambda_i\}}^{\Lambda}\left(x_i,{\bf k}_i^{\perp}\right)\,.
\label{tmdeq}\\
&g^{1L}_{\gamma}(x,{\bf k}^{\perp 2}) = \frac{1}{2}\int [{\rm d}e\bar{e}]\nn\\&\times\sum_{\Lambda,\lambda_i}\lambda_1\,
\Psi_{\{\lambda_i\}}^{\Lambda*}\left(x_i,{\bf k}_i^{\perp}\right)
\Psi_{\{\lambda_i\}}^{\Lambda}\left(x_i,{\bf k}_i^{\perp}\right)\,,
\label{tmdeq_g1l}
\end{align}
respectively, where $\lambda_1=1(-1)$ for the struck parton helicity and we use the abbreviation
\begin{align}
&[{\rm d}e\bar{e}] = \frac{{\rm d}\,x_{e}{\rm d}x_{\bar{e}}\, {\rm d}^2{\bf k}_{e}^{\perp}\,  {\rm d}^2{\bf k}_{\bar{e}}^{\perp}}{2(2\pi)^3} \delta(x_{e} + x_{\bar{e}}-1) \nn\\&\times
\delta^2({\bf k}_{e}^{\perp} + {\bf k}_{\bar{e}}^{\perp}) \delta (x - x_{e}) \delta^2( {\bf k}^{\perp} - {\bf k}^{\perp}_{e})\,.
\end{align}

Although the quark correlation function defining the leading twist TMDs depend on the direction of the transverse momentum, by definition the TMDs themselves only depend on the magnitude of the transverse momentum~\cite{Mulders:1995dh,Bacchetta:2004jz}. 
%The directional dependence of the transverse momentum lie within coefficients that are factored out of the TMDs. Moreover these coefficients for $f^{1}_{\gamma}$ and $g^{1L}_{\gamma}$ are just constant factors~\cite{Kaur:2020emh}

The LFWFs $\Psi_{\{\lambda_i\}}^{\Lambda}\left(x_i,{\bf k}_i^{\perp}\right)$ are boost-invariant and depend only on $x_{i} = k_{i}^+/{P^+}$ and the relative transverse momentum of the constituent electron (positron) of the dressed photon ${\bf k}_{i}^{\perp}$. The physical transverse momentum of the electron (positron) is given by ${\bf p}_{i}^{\perp} = {\bf k}_{i}^{\perp}+x_{i} {\bf P}^{\perp}$ and the longitudinal momentum is $p_{i}^+ = k_{i}^+ = x_{i}P^+$. 

%The BLFQ solutions given in Eq.~(\ref{single}) involve CM excitations. 
We convert the nonperturbative solutions in the single particle coordinates, Eq.~(\ref{single}), to that in the relative coordinates, $\Psi_{\{\lambda_i\}}^{\Lambda}\left(x_i,{\bf k}_i^{\perp}\right)$, by factorizing out the CM motion~\cite{Wiecki:2014ola,Xu:2021wwj,Hu:2020arv}.
%==============================================
\subsection{GPDs} \label{sec:GPDs}
%==============================================
With our LFWFs, the constituent electron (positron) GPDs in the photon are
given by
\begin{align}
&F\left(x,0,t=-{\bf \Delta}^{\perp 2}\right) = \frac{1}{2}\int \{{\rm d}e\bar{e}\}\\
&\times\sum_{\Lambda,\lambda_i}
\Psi_{\{\lambda_i\}}^{\Lambda*}\left(x_i,{\bf k'}_i^{\perp}\right)
\Psi_{\{\lambda_i\}}^{\Lambda}\left(x_i,{\bf k}_i^{\perp}\right)\,,\nn
%\label{tmdeq}
\label{gpdeq_final}\\
&\widetilde{F}\left(x,0,t=-{\bf \Delta}^{\perp 2}\right) = \frac{1}{2}\int \{{\rm d}e\bar{e}\}\\
&\times\sum_{\Lambda,\lambda_i}\lambda_1\,
\Psi_{\{\lambda_i\}}^{\Lambda*}\left(x_i,{\bf k'}_i^{\perp}\right)
\Psi_{\{\lambda_i\}}^{\Lambda}\left(x_i,{\bf k}_i^{\perp}\right)\,,\nn
%\label{tmdeq}
\label{gpdeq_final_helicity}
\end{align}
where for the struck parton i.e., electron, ${\bf k'}_e^{\perp}={\bf k}_e^{\perp}+(1-x_e){\bf \Delta}^\perp$ and for the spectator i.e., positron, ${\bf k'}_{\bar{e}}^{\perp}={\bf k}_{\bar{e}}^{\perp}-(1-x_{\bar{e}}){\bf \Delta}^\perp$ and  the total momentum transferred to the photon is $t=-{\bf \Delta}^{\perp 2}$ and we use 
\begin{align}
\{{\rm d}e\bar{e}\} &= \frac{{\rm d}x_{e}\,{\rm d}x_{\bar{e}}\, {\rm d}^2{\bf k}_{e}^{\perp}  \,{\rm d}^2{\bf k}_{\bar{e}}^{\perp}}{2(2\pi)^3} \delta(x_{e} + x_{\bar{e}}-1) \nn\\
&\times\delta^2({\bf k}_{e}^{\perp} + {\bf k}_{\bar{e}}^{\perp}) \delta (x - x_{e}) \,.
\end{align}
$F(x,0,t)$ and $\widetilde{F}(x,0,t)$ represent the unpolarized and the polarized GPDs, respectively.
Here, we restrict ourselves  to the kinematical region $0<x<1$ at zero skewness. This region corresponds to the situation where the electron (positron) is removed from the initial photon with longitudinal momentum $xP^+$ and reinserted into the final photon with the same longitudinal momentum. Therefore, the momentum transfer occurs purely in the transverse direction. The parton number remains conserved in this kinematical region describing the diagonal $2 \to 2$ overlaps.
%==============================================
%\subsection{Structure functions} \label{sec:structure_function}
%%==============================================
%In the forward limit, $-t=0$, the GPD reduces to the PDF. With the photon PDF, we proceed to calculate the photon structure function using the parton model. Specifically, in the leading order QED, the structure function can be expressed in terms of the PDFs as \cite{Berger:2014rva}%  is related to the forward limit of the GPD  
%\begin{align}
%F_{2,\mathrm{QED}}^{\gamma}(x) &=\sum_{i=e,\bar{e}}e_i^2\,x\,F_i^\gamma(x,0,0)\nn\\&= 2 x\, F_e^\gamma(x,0,0)\,,
%\label{psfeq}
%\end{align}
%where $e_i$ represents the electric charge of the electron (positron). We will compare our BLFQ computations for the photon observables with the perturbative results presented in the following subsection. 
%============================
\section{Perturbative results}
%============================
The LFWFs of the photon can be calculated analytically using perturbation theory. The two particle LFWFs for the photon is given by~\cite{Harindranath:1998pd}
\begin{align}
&\psi_{\lambda_1,\lambda_2}^{\Lambda}\left(
x,{\bf k}^{\perp}
\right) = \frac{1}{M_{\gamma}^2+ \frac{m_e^2+{\bf k}^{\perp 2}}{x(1-x)}}\frac{e}{\sqrt{2(2\pi)^3}}\nn\\
&\times\chi_{\lambda_1}^{\dagger}\Big[
\frac{(\sigma^{\perp}\cdot{\bf k}^{\perp})}{x}\sigma^{\perp} -
 \sigma^{\perp} \frac{(\sigma^{\perp}\cdot{\bf k}^{\perp})}{1-x}\nn\\
&\quad\quad\quad\quad\quad -i \frac{m_e}{x(1-x)}\sigma^{\perp} \Big]\chi_{-\lambda_2}\epsilon_{\Lambda}^{\perp*}\,,
\label{tpwf}
\end{align}
where $x=x_1$ and ${\bf k}^{\perp}={\bf k}_1^{\perp}$ represent the longitudinal momentum fraction and the relative transverse momentum of the electron (positron), and satisfy  $\sum_i x_i =1$ and $\sum_i {\bf k}_i^{\perp} = 0$. 
% and $\sqrt{P^+}\phi_{2}(k_i^+,k_i^{\perp}) = \psi_{2}\left( x_i,k_i^{\perp} \right)$.   The above wavefunction is written in the two component formalism  \cite{Zhang:1993dd,Harindranath:1998pd}. 
The variable $m_e$ is the electron mass, whereas $M_{\gamma}$ is the target mass which we set to zero for the real photon. The helicities for the partons are denoted by $\lambda_i$, while $\Lambda$ denotes the helicity of the photon. Using the LFWFs given in Eq.~(\ref{tpwf}), we compute the perturbative results for the TMDs and the GPDs following Eqs.~(\ref{tmdeq}) and (\ref{gpdeq_final}), respectively and compare with the corresponding results obtained from our BLFQ approach.  
% 
%
%The GPD can be expressed  in terms of the overlap of the two particle light-front wavefunctions as follows:
%\be
%F(x,t) = \int d^2 k^{\perp} ~\psi_{2s_1,s_2}^{*\lambda}\left(
%x,k^{\perp}-(1-x)\Delta^{\perp}
%\right) \psi_{2s_1,s_2}^{*\lambda}\left(
%x,k^{\perp}\right).
%\label{gpdolap}\ee

We obtain the expression for the perturbative result of the photon TMDs as
\begin{align}
&f^{1}_{\gamma \rm (pert)}(x,{\bf k}^{\perp 2}) \nn\\&= \frac{e^2}{8\pi^3} \frac{\left(m_e^2 + {\bf k}^{\perp 2} \left\{x^2+(1-x)^2
\right\}\right)}{\left( {\bf k}^{\perp 2} +M_{\gamma}^2 x(1-x) +m_e^2\right)^2}\,,
\end{align}
\begin{align}
&g^{1L}_{\gamma \rm (pert)}(x,{\bf k}^{\perp 2}) \nn\\&= \frac{e^2}{8\pi^3} \frac{\left(m_e^2 + {\bf k}^{\perp 2} \left\{x^2-(1-x)^2
\right\}\right)}{\left( {\bf k}^{\perp 2} +M_{\gamma}^2 x(1-x) +m_e^2\right)^2}\,.
\end{align}

Meanwhile, the perturbative expression for the unpolarized and helicity dependent GPDs of the photon are given by
\begin{align}
&F_{\rm pert}(x,0,t) = \frac{e^2}{8\pi^3} \Big[
\left\{
x^2+(1-x)^2
\right\}\nn\\&\quad\quad\quad\quad\times(I_1+I_2+\mathcal{L}I_3) +2 m_e^2 I_3
\Big]\,,\\
&\widetilde{F}_{\rm pert}(x,0,t) = \frac{e^2}{8\pi^3} \Big[
\left\{
x^2-(1-x)^2
\right\}\nn\\&\quad\quad\quad\quad\times(I_1+I_2+\mathcal{L}I_3) +2 m_e^2 I_3
\Big]\,,
\end{align}
where 
$
\mathcal{L} = -2m_e^2 + 2M_{\gamma}^2x(1-x) + t(1-x)^2
$
and the integrals $I_i$ are given by
\begin{align}
&I_1 = I_2 = \int \frac{{\rm d}^2 {\bf k}^{\perp}}{D} \nn\\&= \pi \mathrm{Log} \left[
\frac{\Lambda_{\rm UV}^2- M_{\gamma}^2x(1-x)+m_e^2}{\mu^2_{\rm IR} - M_{\gamma}^2x(1-x)+m_e^2}\right] \,, \\
&I_3 = \int \frac{{\rm d}^2 {\bf k}^{\perp}}{DD'} = \int_0^1 d\beta \frac{\pi}{P(x,\beta,t) }\,,
\label{int123}
\end{align}
with $\Lambda_{\rm UV}$ and $\mu_{\rm IR}$ being the UV and IR regulators, respectively, and
%$D = {\bf k}^{\perp 2} - M_{\gamma}^2x(1-x) +m_e^2$, 
%$D' ={\bf k}^{\perp 2} -t (1-x)^2 -2{\bf k}^{\perp}\cdot{\bf\Delta}^{\perp}(1-x)-M_{\gamma}^2x(1-x)+m_e^2$, and  
%$P(x,\beta,t) = -M_{\gamma}^2x(1-x)+m_e^2-t\beta(1-\beta)(1-x)^2$.
\begin{align}
 D &= {\bf k}^{\perp 2} - M_{\gamma}^2x(1-x) +m_e^2\,,\nn \\
D' &={\bf k}^{\perp 2} -t (1-x)^2-M_{\gamma}^2x(1-x) \nn\\
&-2{\bf k}^{\perp}\cdot{\bf\Delta}^{\perp}(1-x) +m_e^2 \,,\nn\\ 
P(x,\beta,t) &= -M_{\gamma}^2x(1-x)+m_e^2\nn\\&-t\beta(1-\beta)(1-x)^2 \,.  
\label{Pint}
\end{align}
%===============================
\section{Numerical results}  \label{secnum}
%===============================
There are three parameters in our BLFQ computation: the electron (positron) mass $m_e$, the  coupling constant of the $\gamma\to e\bar{e}$ vertex $e$, and the 2D HO scale $b$, which we set the same as the mass of the electron, i.e., $b=m_e$. We set the common parameters to both the BLFQ and the perturbative calculations, the electron mass to its physical value, $m_e=0.51$ MeV, and the coupling constant $e=0.3$ that corresponds to the QED fine structure constant $\alpha=g^2/(4\pi)\sim 1/137$. Meanwhile, the mass of the real photon is $M_{\gamma} =0$. Using those model parameters, we then present the photon observables.

The TMDs evaluated in BLFQ have oscillations in the transverse direction owing to the oscillatory behavior of the HO basis functions employed in the transverse plane. The average over the BLFQ results at different $N_{\rm max}$ reduces these finite basis artifacts. Following the averaging procedure reported in Ref.~\cite{Hu:2020arv}, we average the results for three different $N_{\mathrm{max}}$ at fixed $K$. The averaging scheme is followed by taking an average of averages. We first consider the average between the BLFQ computations at $N_{\mathrm{max}} = n$ and $N_{\mathrm{max}} = n + 2$. We take another average of the results obtained at $N_{\mathrm{max}} = n+2$ and $N_{\mathrm{max}} = n + 4$. The final result is then obtained by taking the average between the previous two averages. Therefore, this two-step averaging method involves the BLFQ computations at $N_{\mathrm{max}} = \{n,\, n + 2,\,n+4\}$, while we set $K = n = 100$~\cite{Hu:2020arv}.
%Here again, we find excellent agreement with the perturbative results.

\begin{figure}[ht!]
\begin{center}
 \subfloat[\label{fig:TMDs_a}]{\includegraphics[width=0.75\linewidth]{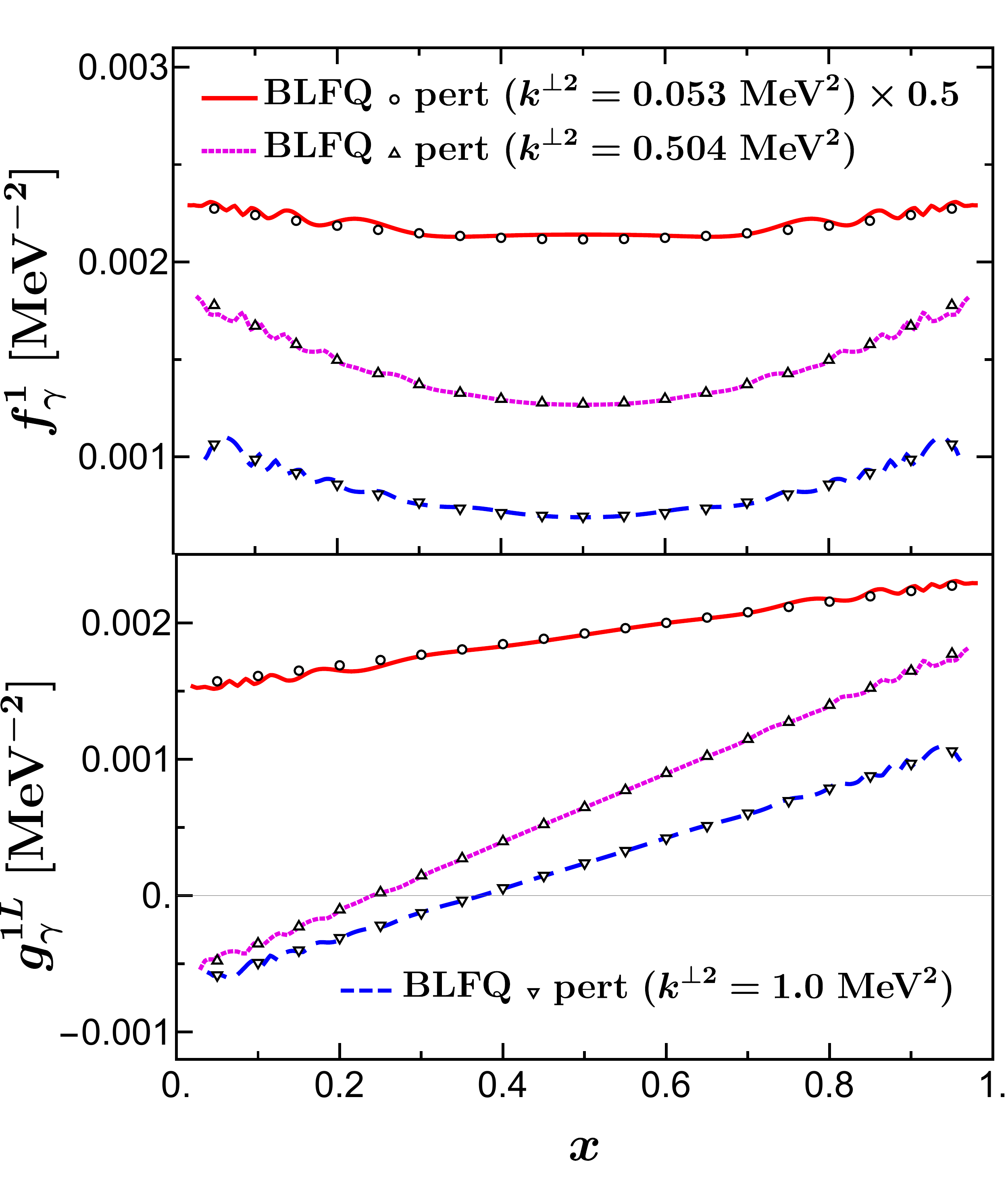}}\\
\subfloat[\label{fig:TMDs_b}]{\includegraphics[width=0.75\linewidth]{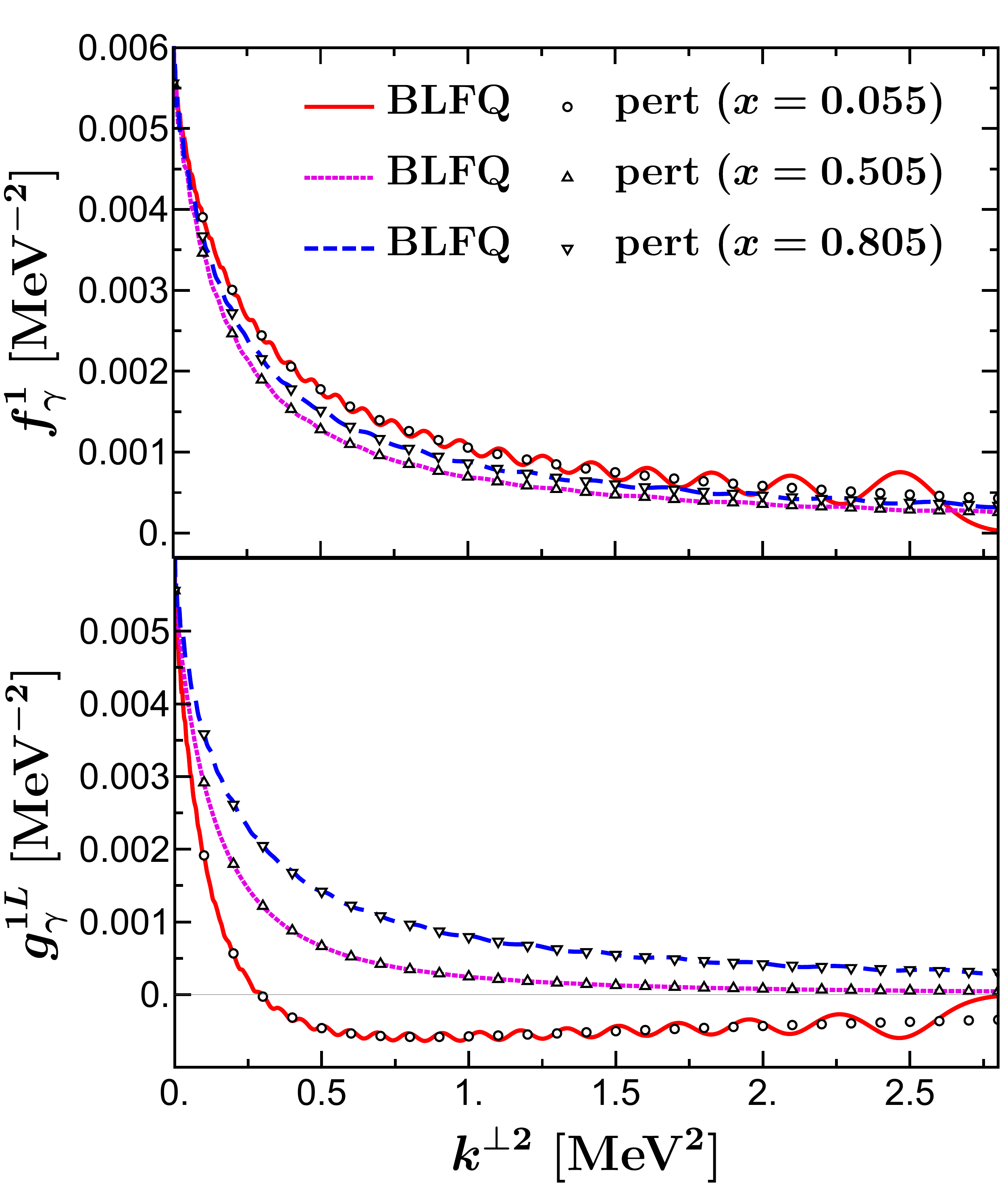}}
\caption{The unpolarized $f_\gamma^1(x,{\bf k}^{\perp 2})$ and helicity $g_\gamma^{1L}(x,{\bf k}^{\perp 2})$ TMDs of the photon. (a) shows the TMDs as functions of $x$ for fixed values of ${\bf k}^{\perp 2}$, whereas (b) displays the TMDs as functions of ${\bf k}^{\perp 2}$ for fixed values of $x$. The BLFQ results (lines) are compared with the perturbative results (circle and triangle symbols). The BLFQ results are obtained by averaging, as described in the text, over the BLFQ computations at $N_{\mathrm{max}} = \{100,\,  102,\, 104\}$ and $K=100$.}
\label{fig:TMDs}
\end{center}
\end{figure}

We show the photon's unpolarized $f^1_{\gamma}(x,{\bf k}^{\perp 2})$ and helicity $g_\gamma^{1L}(x,{\bf k}^{\perp 2})$ TMDs in Fig.~\ref{fig:TMDs}, where
we compare our BLFQ computations with the perturbative results.  We find that the BLFQ results for both the unpolarized and the polarized TMDs are in excellent agreement with the perturbative calculations. For a fixed value of ${\bf k}^{\perp 2}$, the unpolarized TMD $f^1_{\gamma}(x,{\bf k}^{\perp 2})$ approaches to its maximum value when either the electron or the positron carries most of the photon's longitudinal momentum, i.e., near the end points of $x\to \{0,1\}$  (see Fig.~\ref{fig:TMDs_b}, top frame). On the other hand, the TMD finds its minima, when the electron and the positron share exactly equal longitudinal momentum of the photon, i.e, at $x =0.5$ (see Fig.~\ref{fig:TMDs_a}, top frame). Meanwhile, this symmetry is broken for the polarized TMD $g_\gamma^{1L}(x,{\bf k}^{\perp 2})$,  
as can be seen from Fig.~\ref{fig:TMDs_a}, bottom frame. As may be expected, both the TMDs for a fixed value of $x$ show the maxima when the transverse momentum of the electron ${\bf k}^{\perp 2} \to 0$ as observed in Figs.~\ref{fig:TMDs_b}. Note that the averaging strategy employed to reduce the oscillatory behavior has its limitations due to basis truncation. The finite basis artifact becomes more prominent at the endpoints of $x$. 

%================================================================
%=============================================

%================================================================
\begin{figure}[ht!]
\begin{center}
 \subfloat[\label{fig:3DTMDs_a}]{\includegraphics[width=0.75\linewidth]{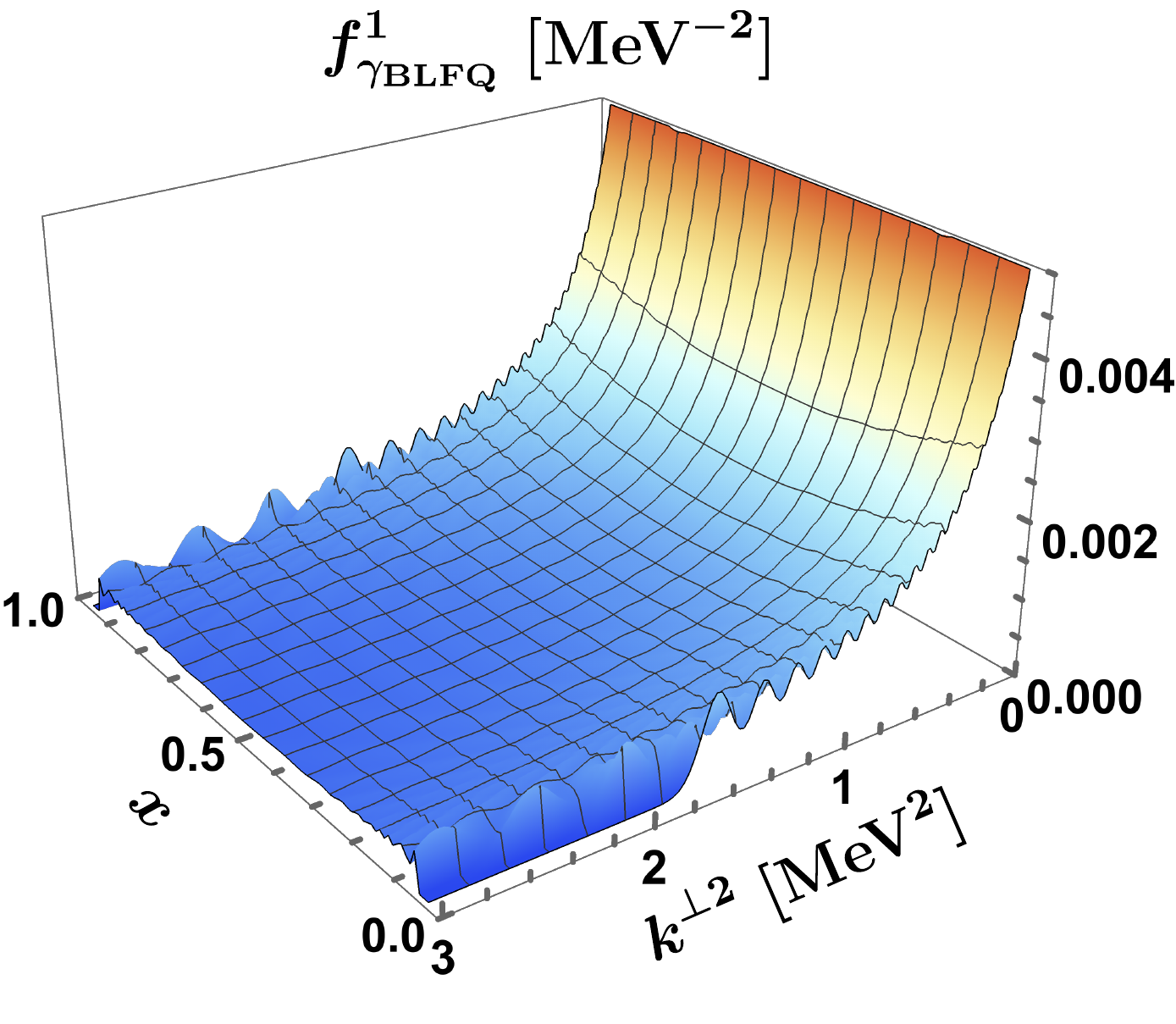}}\\
 \subfloat[\label{fig:3DTMDs_b}]{\includegraphics[width=0.75\linewidth]{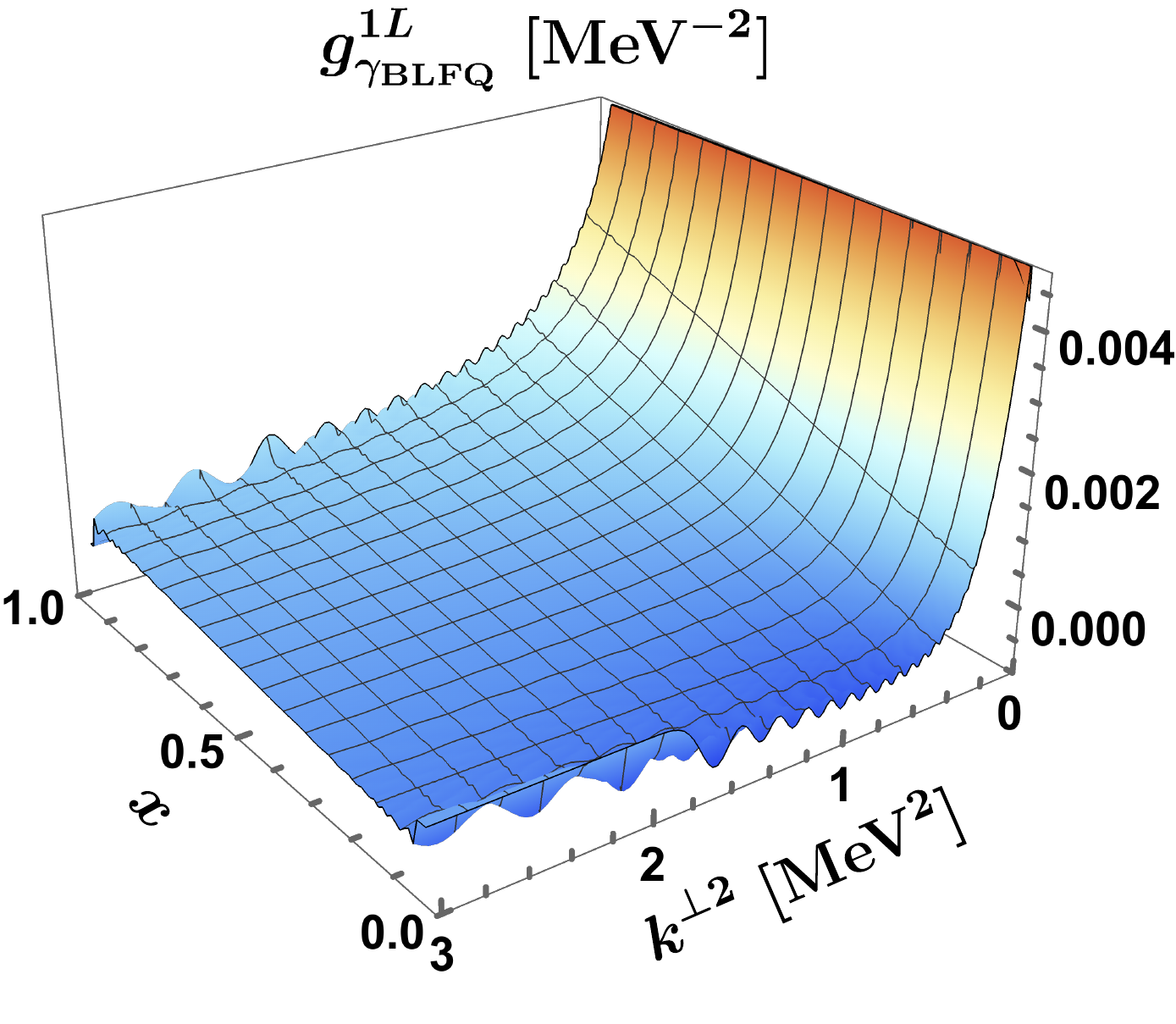}}
\caption{3D plots of the  BLFQ results for (a) $f_\gamma^1(x,{\bf k}^{\perp 2})$ and (b) $g_\gamma^{1L}(x,{\bf k}^{\perp 2})$ TMDs of the photon. The results are obtained by averaging over the BLFQ computations at $N_{\mathrm{max}} = \{100,\,  102,\, 104\}$ and $K=104$.}
\label{fig:3DTMDs}
\end{center}
\end{figure}
%================================================================

The three-dimensional structures of our BLFQ results for the TMDs after averaging are shown in Fig.~\ref{fig:3DTMDs}. It is clearly seen that both TMDs have peaks near ${\bf k}^{\perp 2} \to 0$ and the unpolarized TMD is symmetric around $x=0.5$, while the polarized TMD shows asymmetric behavior with $x$. The peak in the transverse direction in $g_\gamma^{1L}$ falls rapidly to zero or even negative with decreasing $x$. 
%These peaks in $f^1_{\gamma}(x,{\bf k}^{\perp 2})$ and $g_\gamma^{1L}(x,{\bf k}^{\perp 2})$ indicate the dominant probability to find an unpolarized electron and the longitudinally polarized electron, respectively in the physical photon. 

%The three-dimensional structures of our BLFQ result after averaging over the BLFQ computations at $N_{\mathrm{max}} = \{196,\, 198,\, 200\}$, with $K_{\mathrm{max}} = 200$ and the perturbative result for the TMD are compared in Fig.~\ref{fig5}. We observe that the qualitative behavior of the TMD in both the approaches is very similar, while the oscillations appeared in our BLFQ computation are the reflections of the finite basis artifacts.

%=====================================================================

\begin{figure}[ht!]
\begin{center}
 \subfloat[\label{fig:GPDs_a}]{\includegraphics[width=0.75\linewidth]{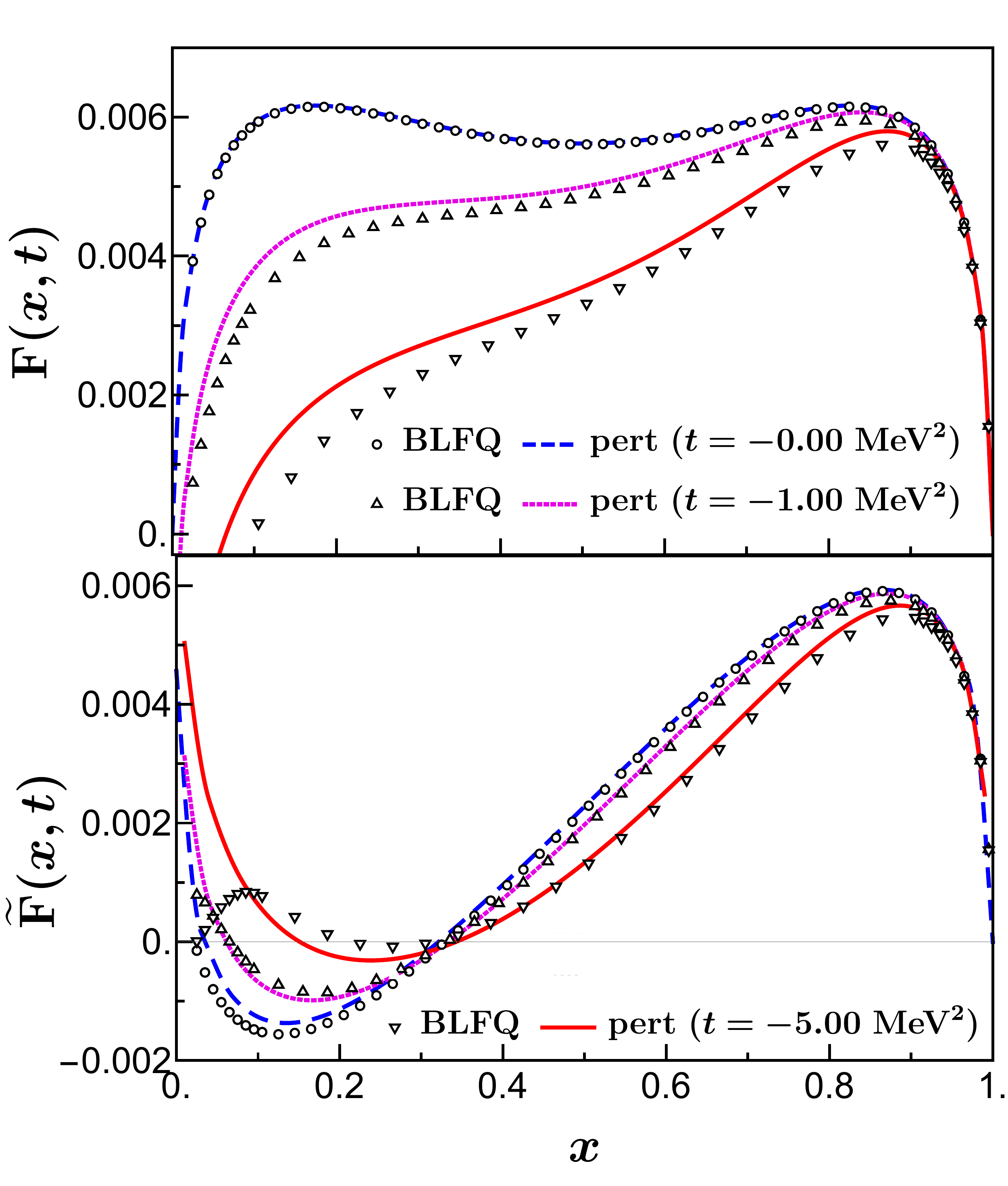}}\\
 \subfloat[\label{fig:GPDs_b}]{\includegraphics[width=0.75\linewidth]{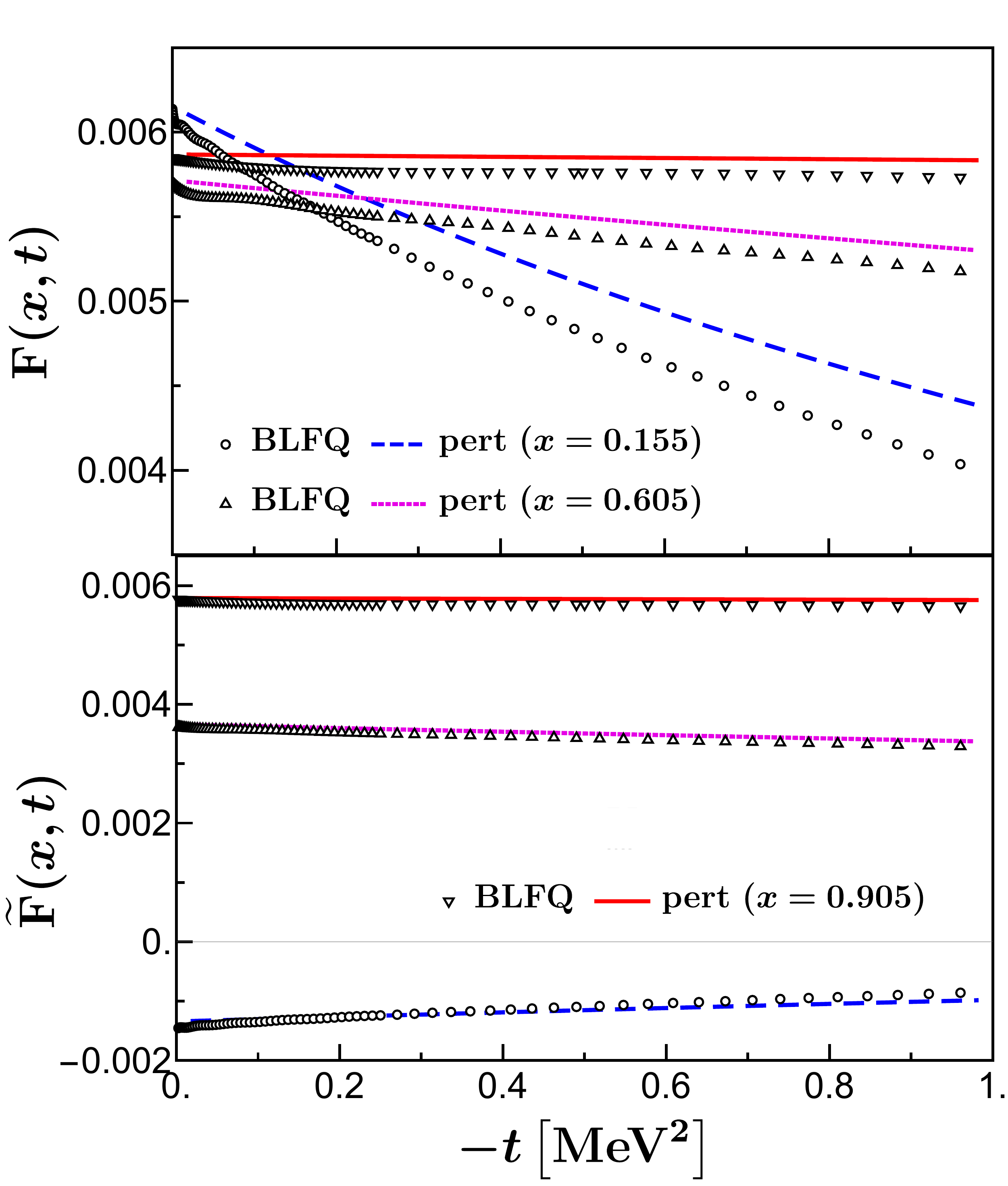}}
\caption{The unpolarized $F(x,t)$ and helicity $\tilde{F}(x,t)$ GPDs of the photon. (a) shows the GPDs as functions of $x$ for fixed values of $-t$, whereas (b) displays the GPDs as functions of $-t$ for fixed values of $x$. The BLFQ results (circle and triangle symbols) are compared with the perturbative results (lines). The BLFQ results are obtained at $N_{\mathrm{max}}=K=100$.}
\label{fig:GPDs}
\end{center}
\end{figure}

Figure~\ref{fig:GPDs} compares the BLFQ results for the electron unpolarized $F(x,t)$ and helicity $\tilde{F}(x,t)$ GPDs in the physical photon at different values of the momentum transfer $-t =\{0,\, 1,\,5\}~ \mathrm{MeV^2}$ with the corresponding perturbative results. Here, we consider the momentum transfer to be purely in the transverse direction. %We obtain these results with at $N_{\rm max}=100$.
% evaluated using the UV cutoffs related to the respective $N_{\mathrm{max}}=100$. 
 The basis truncation parameter $N_{\mathrm{max}}$ acts implicitly as the UV regulator for the LFWFs in the transverse direction, with a UV cutoff $\Lambda_{\rm UV}\approx b\sqrt{x(1-x)2N_{\rm max}}$.  Note that we use the same UV cutoff while computing the perturbative results. Here again, we find good agreement with the perturbative results when the momentum transferred is low. However, with increasing momentum transferred, the BLFQ results at low-$x$ deviate from the perturbative results. % as may be expected from the basis UV cutoff mentioned above.
The deviation at low-$x$ is a direct consequence of the approximate nature of the UV cutoff in our BLFQ framework. This is evident from the relatively large amplitude of the oscillations observed in the TMDs at low-$x$ (see Fig.~\ref{fig:TMDs}). 
%Due to these oscillations, the UV cutoff in BLFQ is not sharp. 
As we increase $N_{\rm max}$, the UV cutoff increases and we expect that our results will slowly converge towards the perturbative results~\cite{Zhao:2014xaa}. 
 %Note that the BLFQ results are obtained at $N_{\rm max}=K=100$ and we expect that with increasing basis size our results will slowly converge towards the perturbative results.
 In order to check this convergence we quantify the difference between our BLFQ results and the perturbative results for the GPDs in Table~\ref{table1}. We define the maximum percentage difference for a fixed value of $-t$ and with $x$ in the closed interval $x_{\mathrm{i}} = \left[x_{\mathrm{min}},x_{\mathrm{max}}\right]$ as 
 $\delta F_{t}\left(x_{\mathrm{i}} \right) = 100 \times \mathrm{Max}\left[\left|\frac{ F_{\mathrm{BLFQ}}\left( x_{\mathrm{i}},t\right)-F_{\mathrm{pert}}\left( x_{\mathrm{i}},t\right)}{F_{\mathrm{pert}}\left( x_{\mathrm{i}},t\right)}\right|\right]$. Similarly $\delta F_{x}\left(t_{\mathrm{i}} \right)$ denotes for a fixed value of $x$ and with $-t$ in the range $t_{\mathrm{i}} = \left[t_{\mathrm{min}},t_{\mathrm{max}}\right]$. We show this maximum difference for three values of 
$N_{\mathrm{max}} = \{50,\, 100,\,150\}$ in Table~\ref{table1} where we observe the convergence in the form of 
decreasing value of the percentage difference with increasing basis truncation parameter $N_{\mathrm{max}}$.
 
%===============================================================
%==========================================
\addtolength{\tabcolsep}{-2pt} 
\begin{table}[ht]
\centering
\begin{tabular}[t]{|l|cc|ccc|}
\hline
$N_{\mathrm{max}}$~&$\delta F_{t}(x_{\mathrm{i}})$&$\delta \widetilde{F}_{t}(x_{\mathrm{i}})$~&~$\delta F_{x}(t_{\mathrm{i}})$~&~$\delta \widetilde{F}_{x}(t_{\mathrm{i}})$& \\
\hline
~$~50$~&$11.5\%$~&$3.9\%$~&~$4.0\%$~&~$4.5\%$&\\
~$100$~&$6.4\%$~&$2.4\%$~&~$2.6\%$~&~$2.7\%$&\\
~$150$~&$4.8\%$~&$1.8\%$~&~$2.0\%$~&~$2.1\%$&\\
%~$~50$~&$11.515\%$~&$3.865\%$~&~$3.984\%$~&~$4.479\%$&\\
%~$100$~&$6.369\%$~&$2.424\%$~&~$2.586\%$~&~$2.720\%$&\\
%~$150$~&$4.761\%$~&$1.814\%$~&~$2.015\%$~&~$2.079\%$&\\
\hline
\end{tabular}
\caption{The maximum percentage difference as defined in the text between our BLFQ result and the perturbative result for the GPDs. The second column is for a fixed value of $-t = 1.0~\mathrm{MeV^2}$ and with $x_{\mathrm{i}} = [0.2,0.8]$. The third column is for a fixed value of $x \approx 0.605$ with $-t_{\mathrm{i}} = [0.0,1.0]~\mathrm{MeV^2}$. $\delta\widetilde{F}$ correspond to the polarized GPD. Here $N_{\mathrm{max}} = K$.}
\label{table1}
\end{table}%
%================================================================

\begin{figure}[ht!]
\begin{center}
 \subfloat[\label{fig:3DGPDs_a}]{\includegraphics[width=0.75\linewidth]{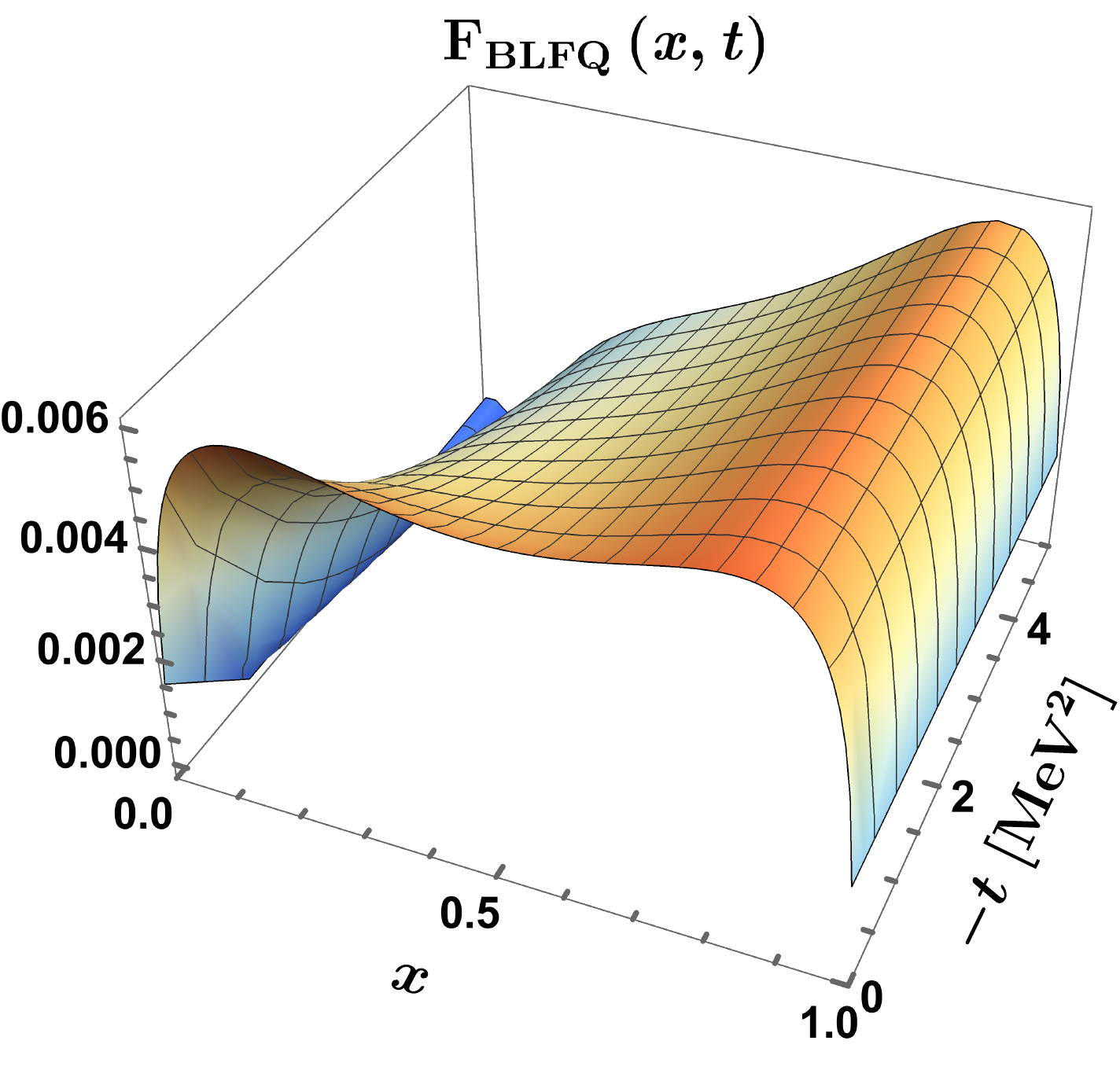}}\\
\subfloat[\label{fig:3DGPDs_b}]{\includegraphics[width=0.75\linewidth]{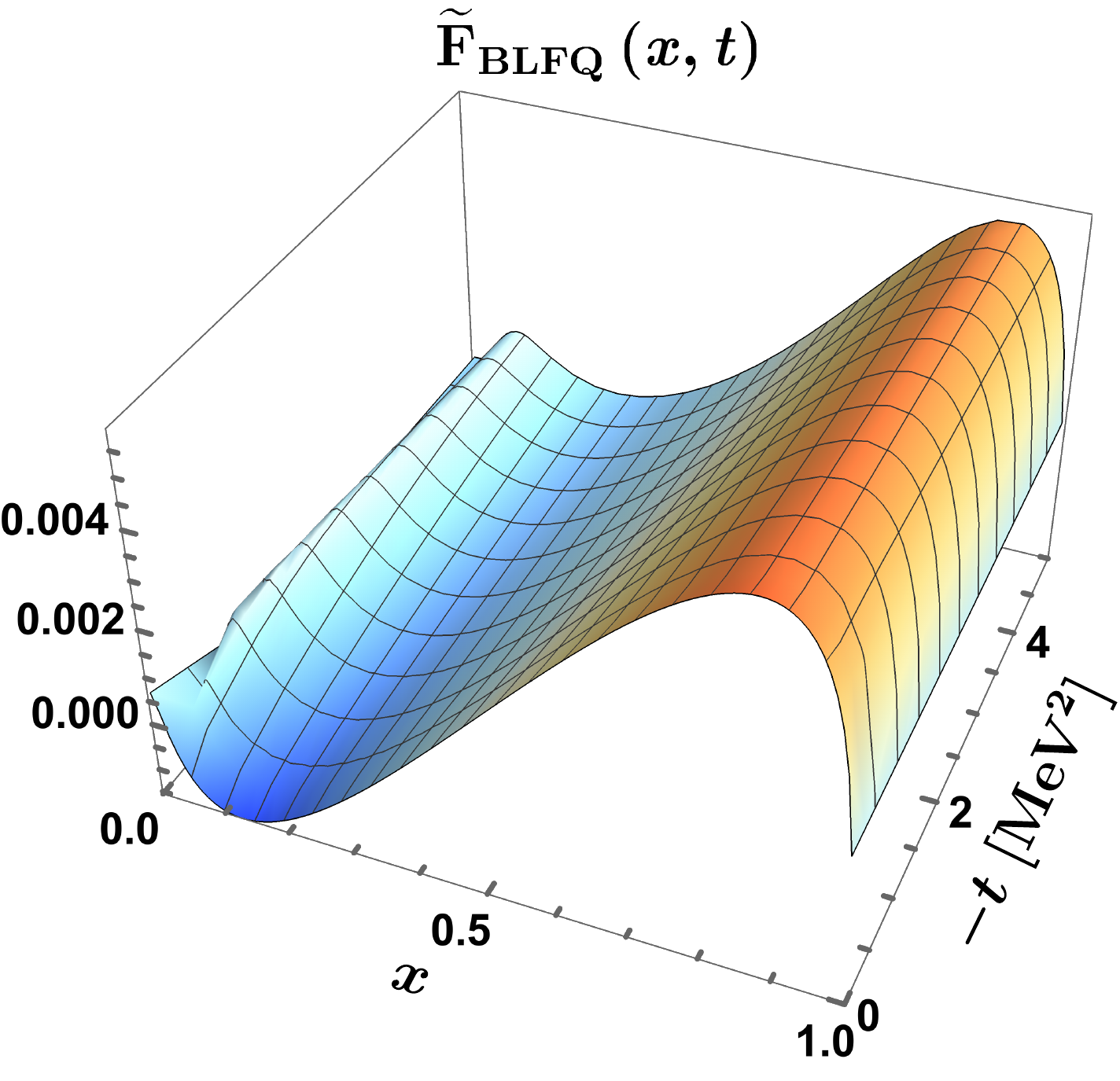}}
\caption{3D plots of the  BLFQ results for (a) $F(x,t)$ and (b) $\widetilde{F}(x,t)$ GPDs of the photon. The results are obtained at $N_{\mathrm{max}}=K=100$.}
\label{fig:3DGPDs}
\end{center}
\end{figure}
%================================================================

In the forward limit ($-t=0$), the unpolarized GPD reduces to the ordinary PDF, which is symmetric over $x=0.5$. This is due to the fact that the electron and positron are equally massive. However, the symmetric nature of the distribution is broken by the momentum transfer. 
We observe that at large-$x$, the photon GPD is nearly independent of $-t$, while at small-$x$, the magnitude of the distribution decreases uniformly as $-t$ increases, as shown in Fig.~\ref{fig:GPDs}. This behavior can be understood from the analytic expression of the perturbative results given in Eq.~(\ref{Pint}). The $-t$ dependence of the GPD comes with the factor $(1-x)^2$, which is responsible for such behavior mentioned earlier. Meanwhile, the polarized GPD shows a minima at low-$x$, whereas it has a maxima at large-$x$ (see Fig.~\ref{fig:GPDs_a}). The magnitude of the valley increases uniformly and its position shifts towards higher-$x$ as the momentum transfer increases, while the peak remains almost the same with varying $-t$. At large-$x$, the polarized GPD is also almost independent of the momentum transfer.  Both the BLFQ and the perturbative calculations exhibit more or less similar features of the GPDs. 

The three-dimensional structure of our BLFQ results for the GPDs are shown in Fig.~\ref{fig:3DGPDs}. Various cuts across these surfaces are depicted in Fig.~\ref{fig:GPDs}.

%=============================================
\section{Conclusion and outlook}
\label{con}
%=============================================
Basis Light-front Quantization has been proposed as a nonperturbative framework for solving quantum ﬁeld theory. In this work, we have applied this approach to QED and study the physical photon in bases truncated to include the $|\gamma\rangle$ and $|e\bar{e}\rangle$ Fock-sectors. We have investigated the unpolarizd and the polarized TMDs and GPDs of the physical photon from its light-front wave functions. These wave functions have been obtained from the eigenvectors of the light-front QED Hamiltonian in the light-cone gauge. 

The BLFQ results have been compared with leading-order perturbative calculations. With a proper renormalization procedure and a rescaling of the naive TMDs and GPDs along with an averaging procedure to minimize the correcting the artifacts introduced by the Fock
space truncation, the BLFQ results are consistent with the perturbative calculations.  These calculations establish a comprehensive and accurate test of the BLFQ approach. The main purpose of this study is to establish the foundation for studying the nonperturbative structure for strongly interacting QCD systems, like vector mesons.

Since the photon LFWFs encode all the information on the photon structure, these can be further employed to evaluate other observables which measure the photon structure in QED, such as the electromagnetic and gravitational form factors, Wigner distributions and spin-orbit correlations etc.
For further investigation, future developments will focus on the virtual photon. With the framework of BLFQ, the LFWFs of the virtual photon can be obtained by renormalizing the bare photon mass to a non-zero value. The virtual photon wave functions describing the $\gamma^*\to$ quark-antiquark splitting can be further employed to investigate the exclusive vector meson production in virtual photon-proton (or nucleus) scattering.~\cite{Lappi:2020ufv,Mantysaari:2021ryb,Shi:2021taf,Ahmady:2016ujw,Dosch:1996ss,Chen:2016dlk}.

%In this work we calculated the photon observables using the formalism of basis light-front quantization. The observables calculated were  the structure functions, TMDs and GPDs. The TMDs were also calculated for the case of a massive photon. The results obtained were compared with leading-order perturbative results. Our calculation also employed the sector dependent renormalization technique and the rescaling of the wavefunction which is required to compensate for any artifacts coming from the Fock sector truncation. We found that our results were in agreement with the perturbative results. Thus this work shows the reliability of the BLFQ approach in solving such bound state problems.

\section*{Acknowledgements}
S. N and C. M. thank the Chinese Academy of Sciences Presidents International Fellowship Initiative for the support via Grants No. 2021PM0021 and 2021PM0023, respectively. C. M. is supported by new faculty start up funding by the Institute of Modern Physics, Chinese Academy of Sciences, Grant No. E129952YR0.  X. Z. is supported by new faculty startup funding by the Institute of Modern Physics, Chinese Academy of Sciences, by Key Research Program of Frontier Sciences, Chinese Academy of Sciences, Grant No. ZDB-SLY-7020, by the Natural Science Foundation of Gansu Province, China, Grant No. 20JR10RA067 and by the Strategic Priority Research Program of the Chinese Academy of Sciences, Grant No. XDB34000000. A. M. thanks SERB for support through POWER Fellowship (SPF/2021/000102). J. P. V. is supported in part by the Department of Energy under Grants No. DE-FG02-87ER40371, and No. DE-SC0018223 (SciDAC4/NUCLEI). 
%This research used resources of the National Energy Research Scientific Computing Center (NERSC), a U.S. Department of Energy Office of Science User Facility operated under Contract No. DE-AC02-05CH11231. 
A portion of the computational resources were also provided by Gansu Computing Center.

%==============================================

\end{document}